\documentclass[12pt]{article}
\usepackage{epsfig}

\newcommand{\mysection}{\setcounter{equation}{0}\section}

\def\beq{\begin{equation}}
\def\eeq{\end{equation}}
\def\beqa{\begin{eqnarray}}
\def\eeqa{\end{eqnarray}}

\newlength{\dinwidth} \newlength{\dinmargin}
\begin{document}

\begin{center}
{\Large \bf Top-quark transverse-momentum distributions in $t$-channel single-top production}
\end{center}
\vspace{2mm}
\begin{center}
{\large Nikolaos Kidonakis}\\
\vspace{2mm}
{\it Kennesaw State University,  Physics \#1202,\\
1000 Chastain Rd., Kennesaw, GA 30144-5591, USA} \\
\end{center}

\begin{abstract}
I present approximate next-to-next-to-leading-order (NNLO) top-quark 
transverse momentum, $p_T$, distributions in $t$-channel single-top production. 
These distributions are derived from next-to-next-to-leading-logarithm (NNLL) 
soft-gluon resummation. Theoretical results for the single top as well as
 the single antitop $p_T$ distributions are shown for LHC and Tevatron energies. 
\end{abstract}

\mysection{Introduction}

Single-top production has been observed at both the Tevatron \cite{D0st,CDFst} 
and the LHC \cite{CMStch,ATLAStch} and it has been an important process for 
study in addition to top-antitop pair production. 
The single-top cross sections are smaller than the corresponding 
ones for top-pair production and thus more difficult to observe. A lot of 
theoretical progress has been made in calculating the total cross sections and 
differential distributions.

Single-top production can proceed via three different types of  
partonic processes. One of them is the $t$-channel process via
the exchange of a space-like $W$ boson, a second is the $s$-channel 
via the exchange of a time-like $W$ boson, 
and a third is associated $tW$ production. 
At both LHC and Tevatron energies the $t$-channel is numerically dominant.
The $t$-channel partonic processes are of the form $qb \rightarrow q' t$ 
and ${\bar q} b \rightarrow {\bar q}' t$ for single top production, and 
$q {\bar b} \rightarrow q' {\bar t}$ 
and ${\bar q} {\bar b} \rightarrow {\bar q}' {\bar t}$ 
for single antitop production.

The calculation of the complete next-to-leading order (NLO) corrections to the 
differential cross section for $t$-channel production was performed in 
Ref. \cite{BWHL}. This calculation enabled the derivation of the top quark $p_T$ distribution at NLO. More recent results and further studies for the NLO top $p_T$ distribution in $t$-channel production have appeared in \cite{CFMT,SYMC,FGMS,FRT,PF}. 

Theoretical calculations for $t$-channel production 
beyond NLO that include higher-order corrections 
from next-to-leading-logarithm (NLL) soft-gluon resummation 
appeared in \cite{NKsingletopTev,NKsingletopLHC}, and more recently  
at next-to-next-to-leading-logarithm (NNLL) accuracy in \cite{NKtch}.
It was shown in those papers that the soft-gluon corrections dominate the cross 
section at NLO, and thus approximate it very well, while the NNLO soft-gluon 
corrections provide an additional enhancement.

The work in \cite{NKsingletopTev,NKsingletopLHC,NKtch} is at the 
double-differential level and thus allows the calculation not only of 
total cross sections but also of differential distributions. The transverse 
momentum, $p_T$, distribution of the top quark (or the antitop quark) is 
particularly interesting since deviations from new physics may appear at 
large $p_T$, and measurements of the $p_T$ distribution are taken at the 
LHC. 
The calculation of these distributions at LHC as well as Tevatron 
energies is the subject of this paper. Some preliminary results based on the 
work in this paper have appeared in \cite{NKconf}. The work presented here is based on the formalism of the standard moment-space perturbative QCD resummation of soft-gluon corrections. 
Results based on another approach, soft-collinear effective theory (SCET), have also recently appeared in \cite{WLZ}. The differences between the moment-space and SCET approaches to resummation have been detailed in \cite{NKBP}.

\mysection{Top quark $p_T$ distributions}

\begin{figure}
\begin{center}
\includegraphics[width=8cm]{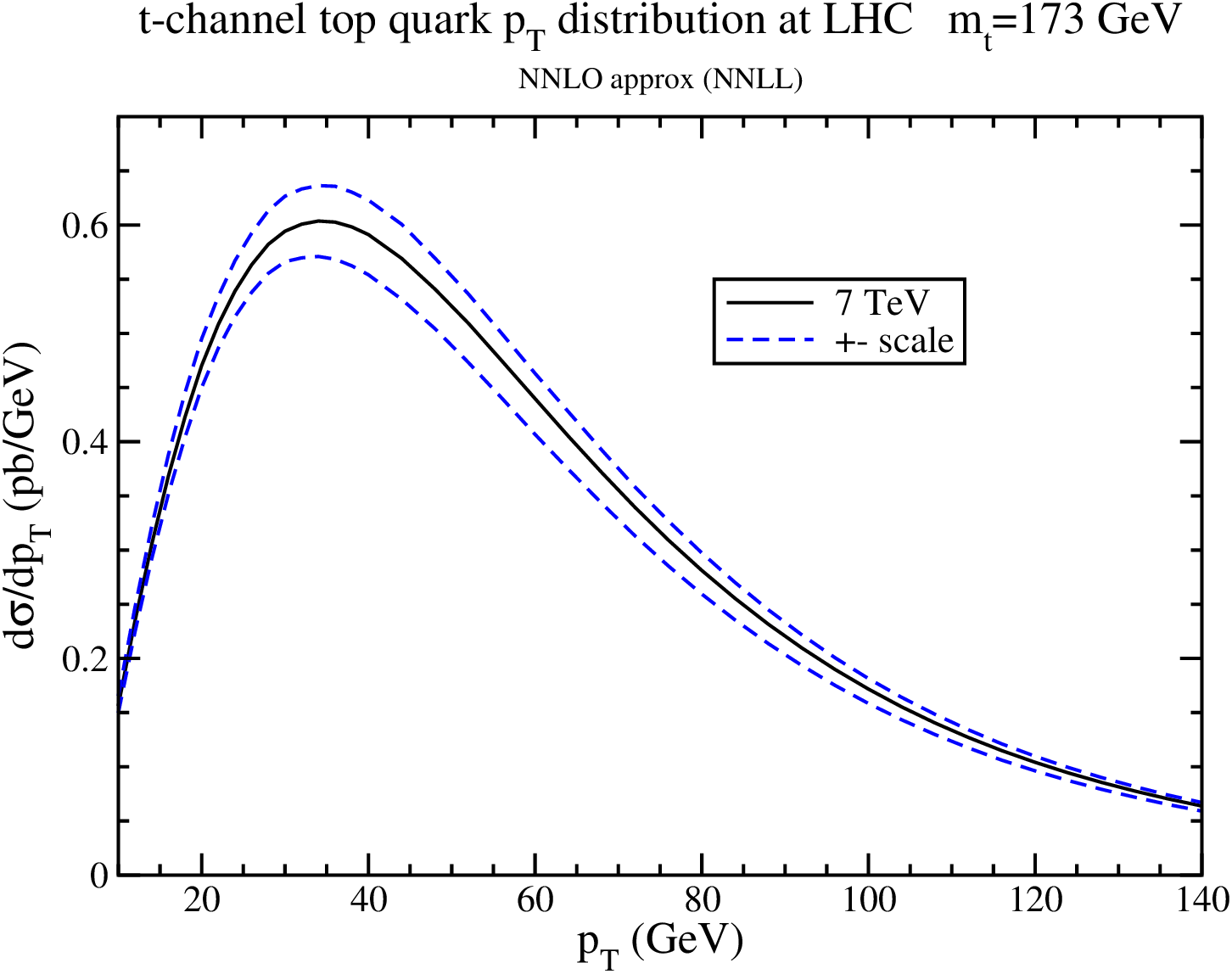}
\hspace{3mm}
\includegraphics[width=8cm]{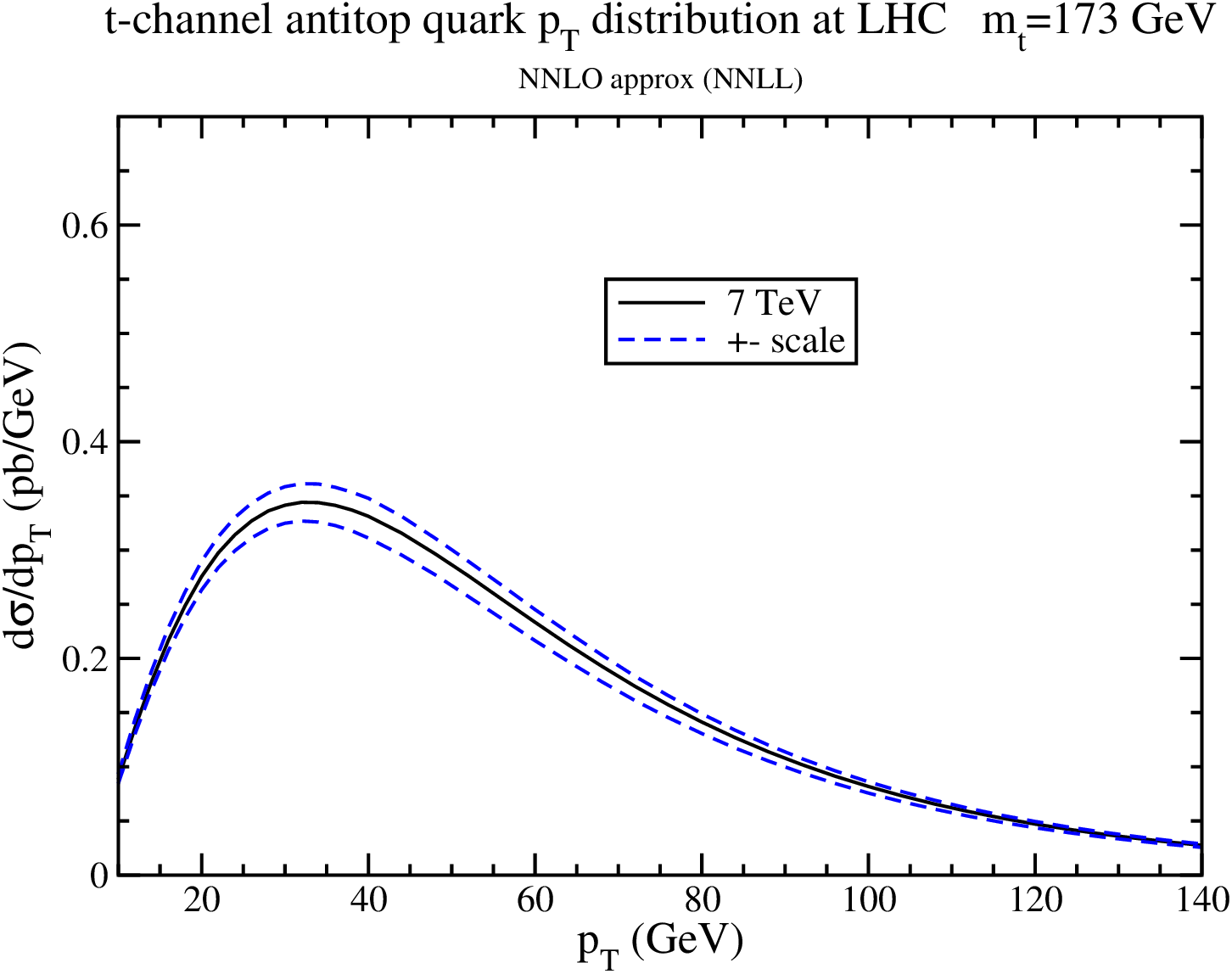}
\caption{NNLO approximate top-quark (left) and antitop (right) 
$p_T$ distributions at 7 TeV energy at the LHC. 
The central result is with $\mu=m_t$ and the 
uncertainty due to scale variation is displayed.}
\label{pttch7lhcplot}
\end{center}
\end{figure}

We consider single-top production in collisions of hadrons $h_1$ and $h_2$ 
with momenta $p_{h1}+p_{h2} \rightarrow p_3+p_4$, 
and let $p_T$ and $Y$ represent the transverse momentum and rapidity 
of the top quark (or antitop).
The underlying partonic reactions have momenta $p_1+p_2 \rightarrow p_3+p_4$.
The partonic invariants are 
$s=(p_1+p_2)^2$, $t=(p_1-p_3)^2$, $u=(p_2-p_3)^2$, $s_4=s+t+u-m_t^2$, where 
$m_t$ is the top quark mass.
The hadronic invariants are 
$S=(p_{h1}+p_{h2})^2$, $T=(p_{h1}-p_3)^2$, and $U=(p_{h2}-p_3)^2$.

The resummation of threshold logarithms is carried out in moment space and it 
follows from the factorization of the differential cross section into hard, soft, and 
jet functions that describe, respectively, the hard scattering, noncollinear 
soft gluon emission, and collinear gluon emission from the initial- and 
final-state quarks and gluons \cite{NKsingletopTev,NKtch}. The resummed result can then be used as a generator of approximate higher-order corrections and inverted back to momentum space without need for any prescriptions. 
The threshold corrections that arise from soft-gluon emission take the form of logarithmic plus distributions, $[\ln^k(s_4/m_t^2)/s_4]_+$, where $k \le 2n-1$ for the $n$-th order QCD corrections. At NNLO these corrections to the double-differential partonic cross section $d^2{\hat \sigma}/(dt \,du)$ take the form 
\beqa 
\frac{d^2{\hat \sigma}^{(2)}}{dt \, du}&=&F^B \frac{\alpha_s^2(\mu_R^2)}{\pi^2} 
\left\{C_3^{(2)} \left[\frac{\ln^3(s_4/m_t^2)}{s_4}\right]_+
+C_2^{(2)}  \left[\frac{\ln^2(s_4/m_t^2)}{s_4}\right]_+
+C_1^{(2)}  \left[\frac{\ln(s_4/m_t^2)}{s_4}\right]_+
+C_0^{(2)}  \left[\frac{1}{s_4}\right]_+ \right\}
\nonumber \\ &&
\eeqa
where $\alpha_s$ is the strong coupling, $\mu_R$ is the renormalization scale, 
and $F^B$ denotes the Born-level contributions. 
The coefficients $C_i^{(2)}$ are in general functions of $s$, $t$, $u$, $m_t$, $\mu_R$, and the factorization scale $\mu_F$; these coefficients have been determined from two-loop calculations and NNLL resummation for all partonic processes contributing to this channel in \cite{NKtch}. 

To calculate the hadronic differential cross section one has to convolute the partonic result with parton distribution functions (pdf). The dominant partonic processes  
are $ub \rightarrow dt$ and ${\bar d} b \rightarrow {\bar u} t$. 
Additional processes involving only quarks are $cb \rightarrow st$  
and the Cabibbo-suppressed $ub \rightarrow st$, 
$cb \rightarrow dt$ and $us \rightarrow dt$; 
the contributions from even more suppressed processes 
($ub \rightarrow bt$, $cb \rightarrow bt$, $ud \rightarrow dt$, etc.) 
are negligible. Additional processes involving antiquarks and quarks 
are ${\bar s} b \rightarrow {\bar c}t$ and the Cabibbo-suppressed 
${\bar d}b \rightarrow {\bar c}t$, 
${\bar s}b \rightarrow {\bar u}t$  and 
${\bar d} s \rightarrow {\bar u} t$; the contributions from 
even more suppressed processes (${\bar s}s \rightarrow {\bar c}t$,
${\bar d}d \rightarrow {\bar u}t$, ${\bar s}d \rightarrow {\bar c}t$, etc.) 
are negligible.
We use the MSTW2008 NNLO pdf \cite{MSTW} in our numerical results below. 

For the $t$-channel processes of the form $qb \rightarrow q't$  the Born terms are
\beq
F^B_{qb \rightarrow q't}= 
\frac{\pi \alpha^2 V_{tb}^2 V_{qq'}^2}{\sin^4\theta_W}
\frac{(s-m_t^2)}{4s(t-m_W^2)^2} \, .
\eeq
For the $t$-channel processes of the form ${\bar q}b \rightarrow {\bar q'} t$ we have 
\beq
F_{{\bar q}b \rightarrow {\bar q'}t}= 
\frac{\pi \alpha^2 V_{tb}^2 V_{{\bar q}{\bar q'}}^2}{\sin^4\theta_W}
\frac{[(s+t)^2-(s+t)m_t^2]}{4s^2(t-m_W^2)^2} \, .
\eeq
Here $\alpha=e^2/(4\pi)$, $V_{ij}$ denote elements of the CKM matrix, and $\theta_W$ is the Weinberg angle.
The processes and results for single antitop production are entirely analogous to those for single top.

The transverse momentum distribution of the top quark (or antitop) is given by
\beqa
\frac{d\sigma}{dp_T}&=&
2 \, p_T \int_{Y^-}^{Y^+} dY 
\int_{x_2^-}^1 dx_2 
\int_0^{s_{4max}} ds_4 \, 
\frac{x_1 x_2 \, S}{x_2 S+T} \,
\phi(x_1) \, \phi(x_2) \, 
\frac{d^2{\hat\sigma}}{dt \, du}
\nonumber \\ &&
\eeqa
where $\phi$ denote the pdf, 
\beq
x_1=\frac{s_4+m_t^2-x_2U}{x_2 S+T} 
\eeq
with $T=-\sqrt{S} \, p_T \, e^{-Y}$ and $U=-\sqrt{S} \, p_T \, e^{Y}$, 
\beq
Y^{\pm}=\pm \frac{1}{2} \ln \frac{1+\sqrt{1-\frac{4p_T^2}
{S[1-m_t^2/S]^2}}}
{1-\sqrt{1-\frac{4p_T^2}{S[1-m_t^2/S]^2}}}
\eeq
\beq
x_2^-=\frac{m_t^2-T}{S+U}
\eeq
and
\beq
s_{4max}=x_2(S+U)+T-m_t^2 \, .
\eeq
Note that the total cross section can easily be obtained by integrating 
the distribution over $p_T$ from 0 to  
$p_{T \, max}=(S-m_t^2)/(2\sqrt{S})$, and we have checked that we recover 
the total cross section result of \cite{NKtch} which is also in very good 
agreement with both LHC \cite{CMStch,ATLAStch} and Tevatron \cite{D0st,CDFst} 
data.  

In Fig. \ref{pttch7lhcplot} we present the approximate NNLO top-quark $p_T$
distribution in the left plot as well as the approximate NNLO antitop $p_T$
distribution in the right plot at the LHC at 7 TeV energy. The horizontal and vertical scales in the two plots are chosen the same for easier comparison of the relative magnitude of the top versus the antitop distributions. In both cases the central result is with a choice of factorization and renormalization scale equal to the top quark mass, taken as $m_t=173$ GeV, and the theoretical uncertainty from the variation of the scales by a factor of two (i.e. from $m_t/2$ to $2 m_t$) is also shown. The distributions peak at a $p_T$ of around 35 GeV and quickly fall with increasing $p_T$, which is shown up to 140 GeV. 

\begin{figure}
\begin{center}
\includegraphics[width=11cm]{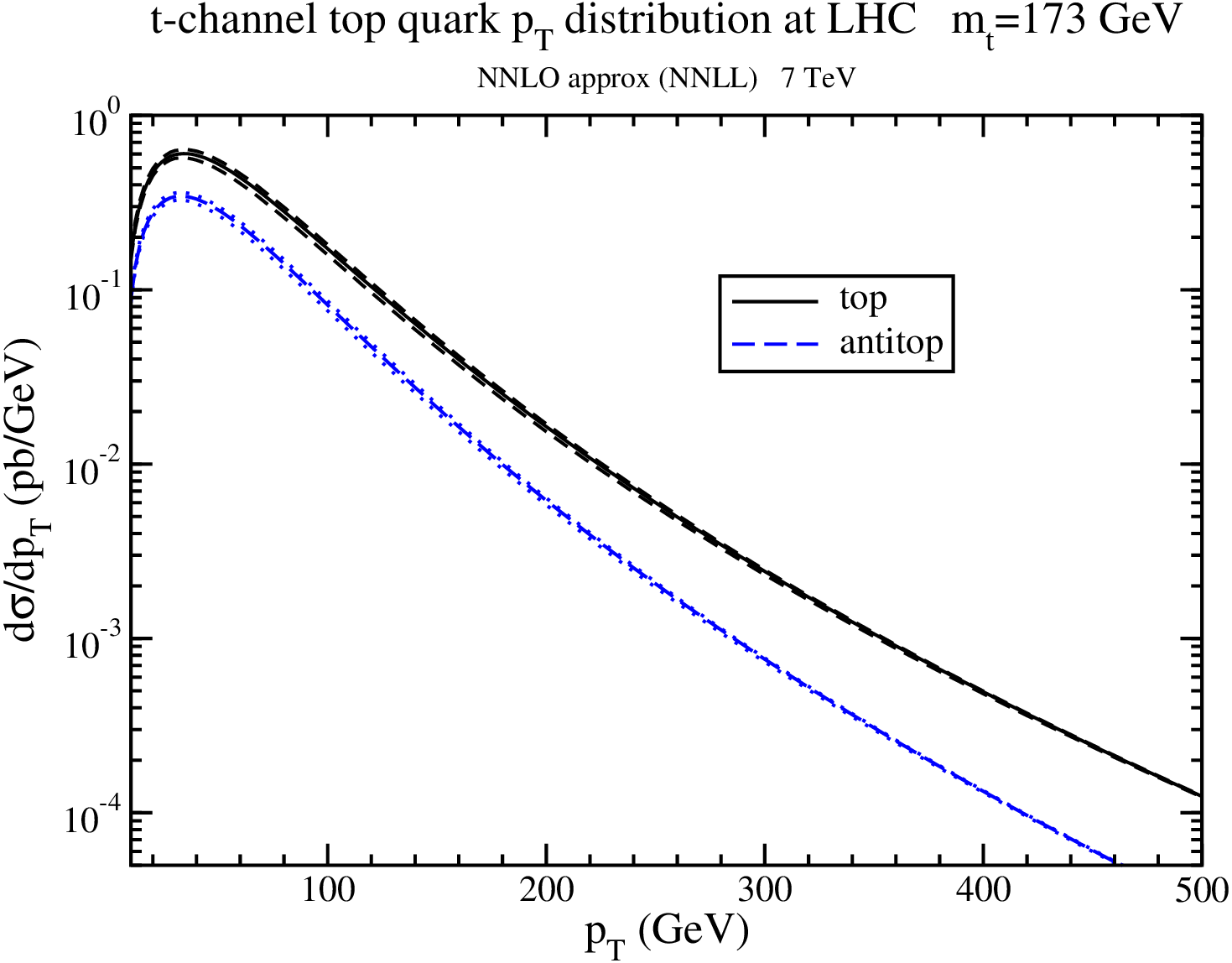}
\caption{NNLO approximate top and antitop  
$p_T$ distributions at 7 TeV energy at the LHC with $p_T$ up to 500 GeV. 
The central results are with $\mu=m_t$ and the 
uncertainty due to scale variation is displayed.}
\label{pttch7lhclogplot}
\end{center}
\end{figure}

In Fig. \ref{pttch7lhclogplot} we present the same distributions at 7 TeV LHC 
energy but now in a logarithmic plot with a much larger $p_T$ range up to 500 
GeV. Here we display all results in one plot for ease of comparison of the top 
versus the antitop $p_T$ distributions. The distributions fall over four orders of magnitude in the $p_T$ range shown.

\begin{figure}
\begin{center}
\includegraphics[width=8cm]{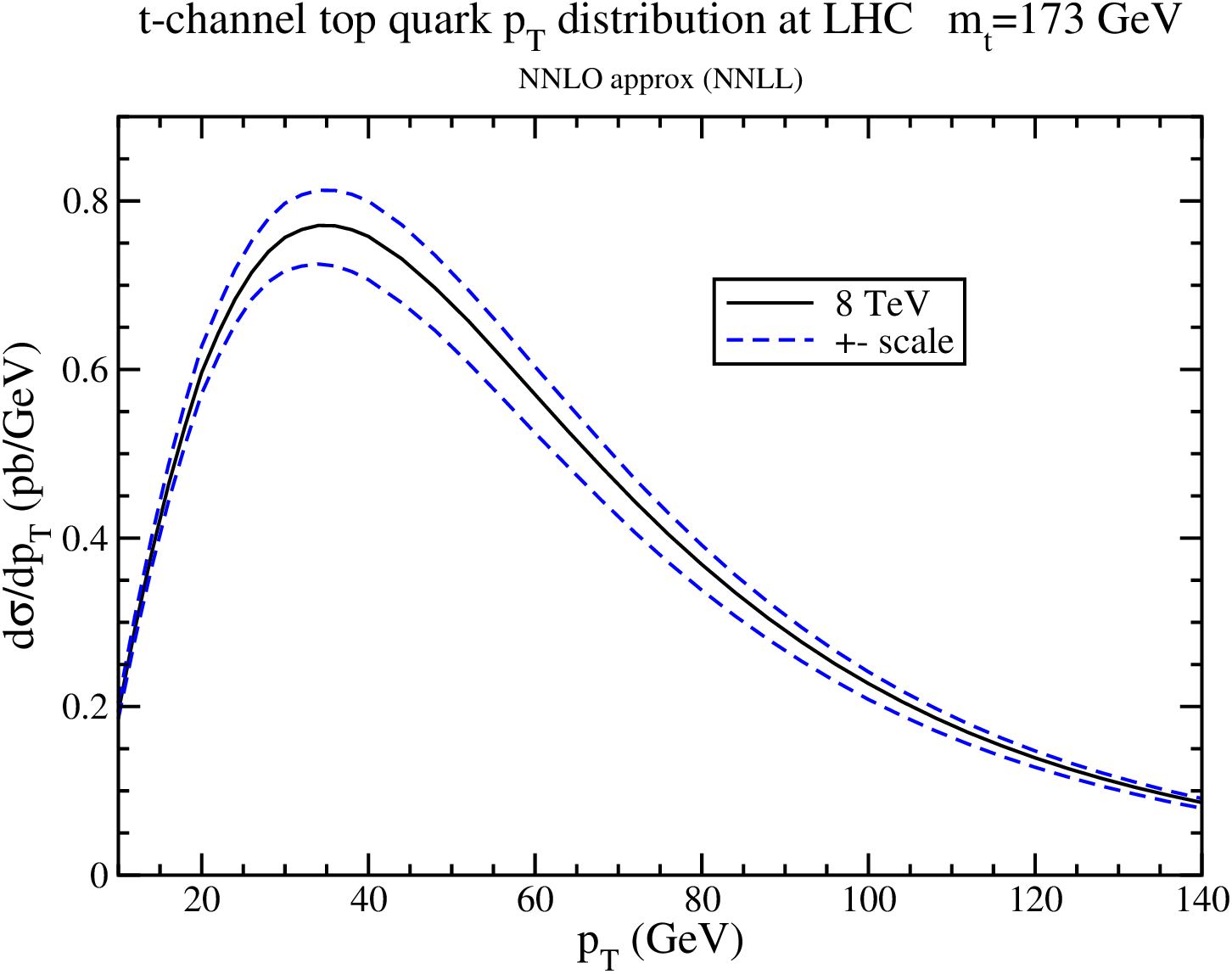}
\hspace{3mm}
\includegraphics[width=8cm]{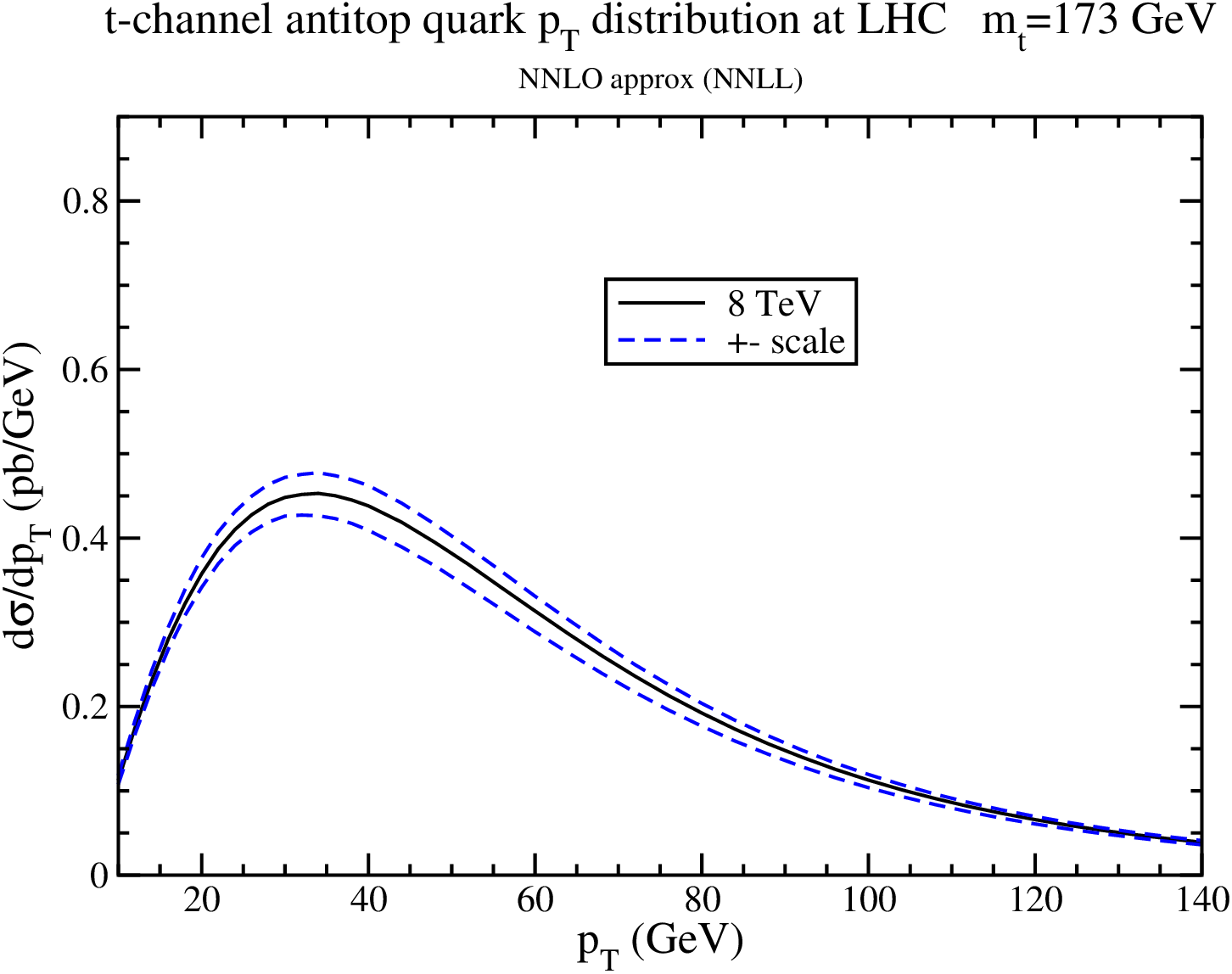}
\caption{NNLO approximate top-quark (left) and antitop (right) 
$p_T$ distributions at 8 TeV energy at the LHC. 
The central result is with $\mu=m_t$ and the 
uncertainty due to scale variation is displayed.}
\label{pttch8lhcplot}
\end{center}
\end{figure}

In Fig. \ref{pttch8lhcplot} we present the approximate NNLO top-quark $p_T$
distribution in the left plot as well as the approximate NNLO antitop $p_T$
distribution in the right plot at the LHC at 8 TeV energy, in analogy to 
Fig. \ref{pttch7lhcplot}. Again the central result is with scales equal to the top quark mass, and the theoretical uncertainty from the variation of the scales by a factor of two is also shown. The distributions again peak at a $p_T$ of around 35 GeV as at 7 TeV energy, with an enhancement over the NLO result of 5\%. 

\begin{figure}
\begin{center}
\includegraphics[width=11cm]{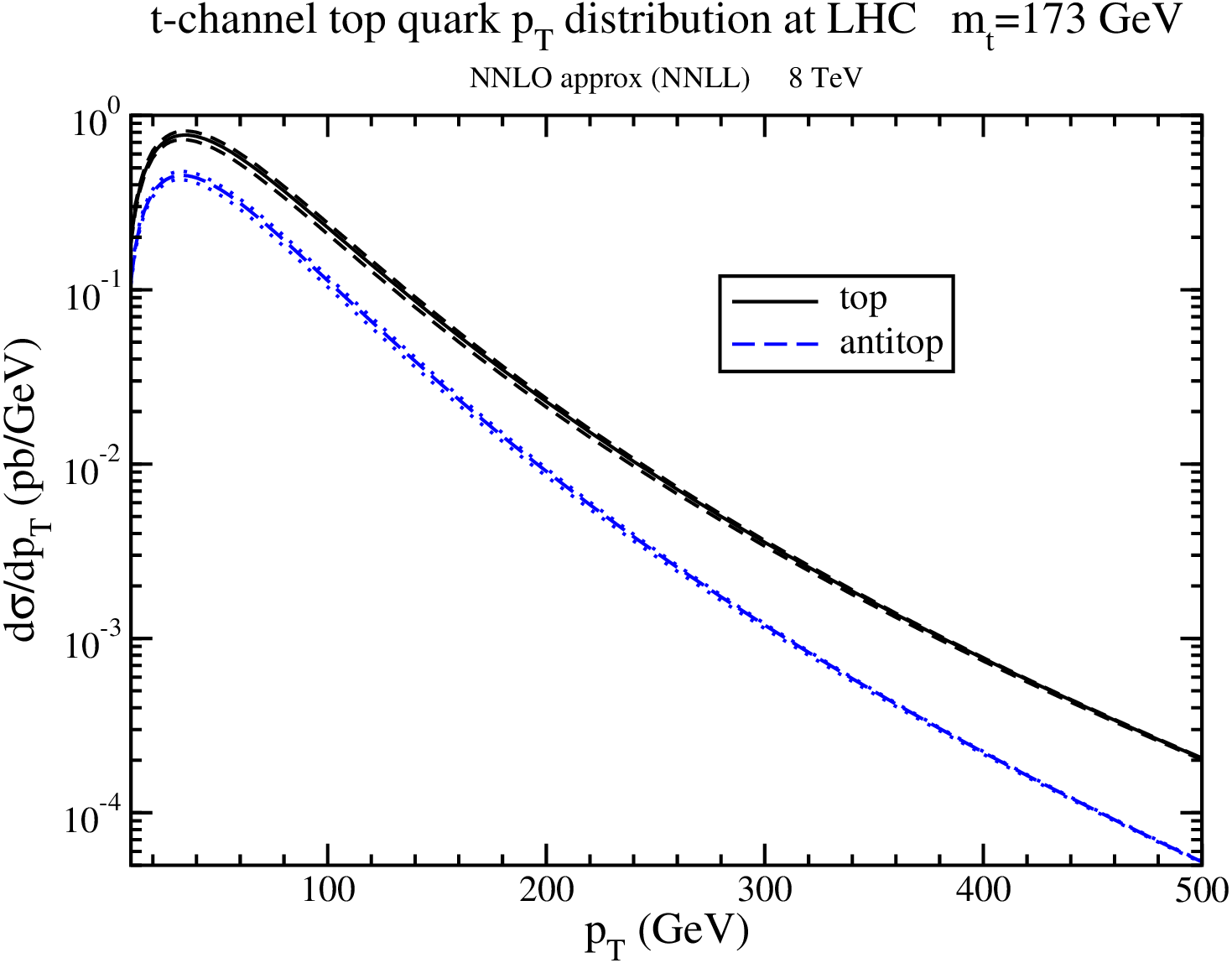}
\caption{NNLO approximate top and antitop 
$p_T$ distributions at 8 TeV energy at the LHC with $p_T$ up to 500 GeV. 
The central results are with $\mu=m_t$ and the 
uncertainty due to scale variation is displayed.}
\label{pttch8lhclogplot}
\end{center}
\end{figure}

In Fig. \ref{pttch8lhclogplot} we present the same distributions at 8 TeV LHC 
energy in one logarithmic plot with a $p_T$ range up to 500 GeV.

\begin{figure}
\begin{center}
\includegraphics[width=8cm]{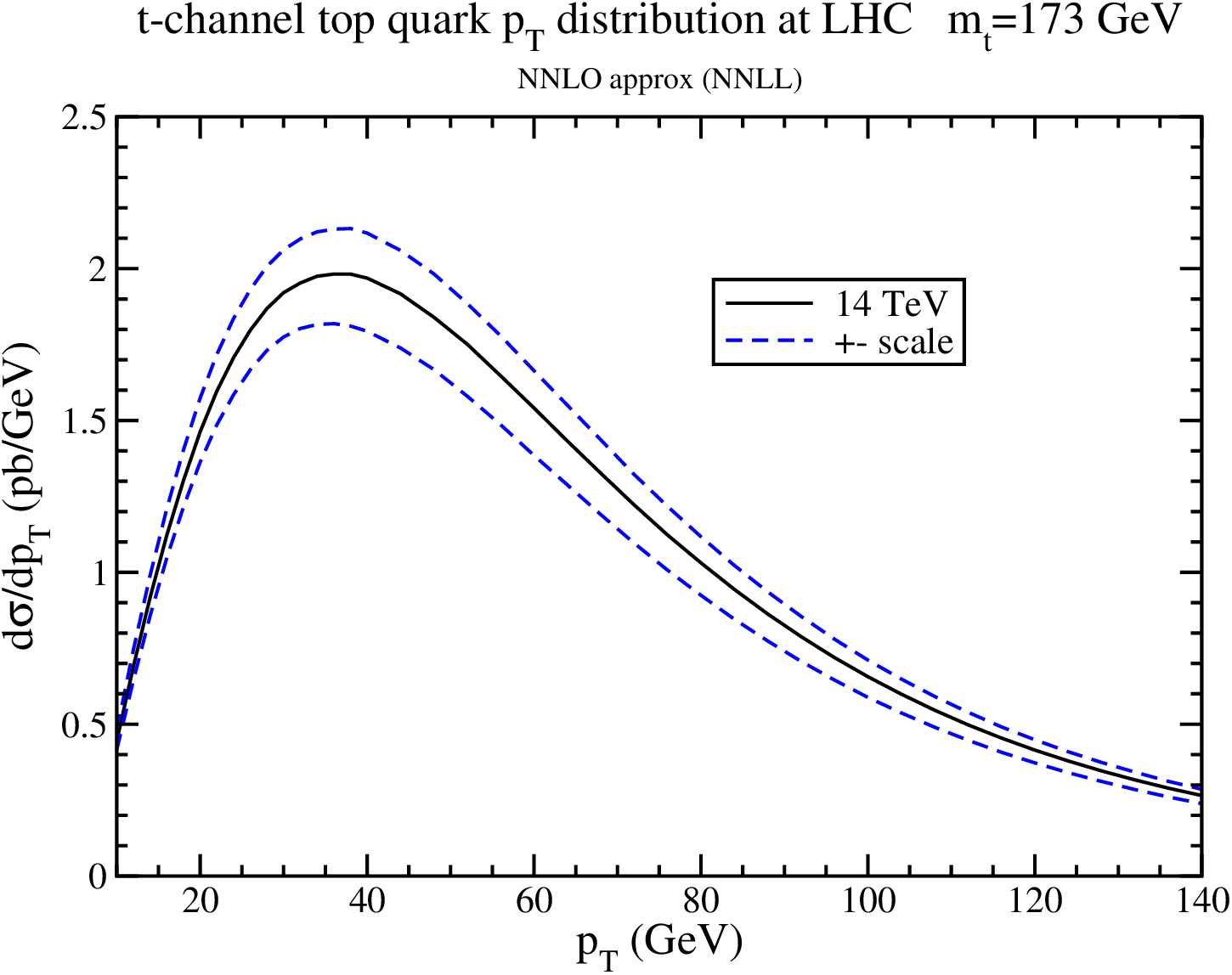}
\hspace{3mm}
\includegraphics[width=8cm]{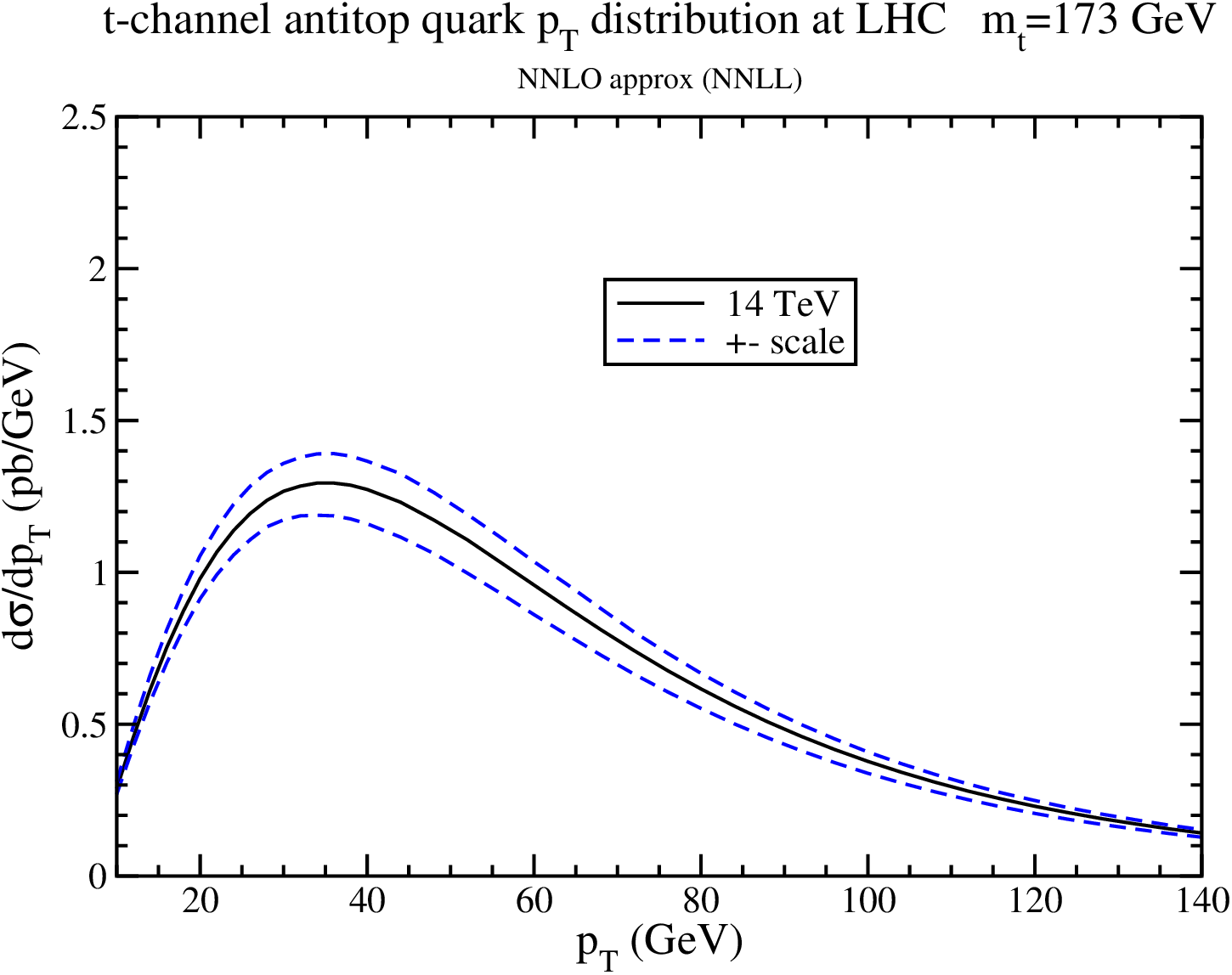}
\caption{NNLO approximate top-quark (left) and antitop (right) 
$p_T$ distributions at 14 TeV energy at the LHC. 
The central result is with $\mu=m_t$ and the 
uncertainty due to scale variation is displayed.}
\label{pttch14lhcplot}
\end{center}
\end{figure}

\begin{figure}
\begin{center}
\includegraphics[width=11cm]{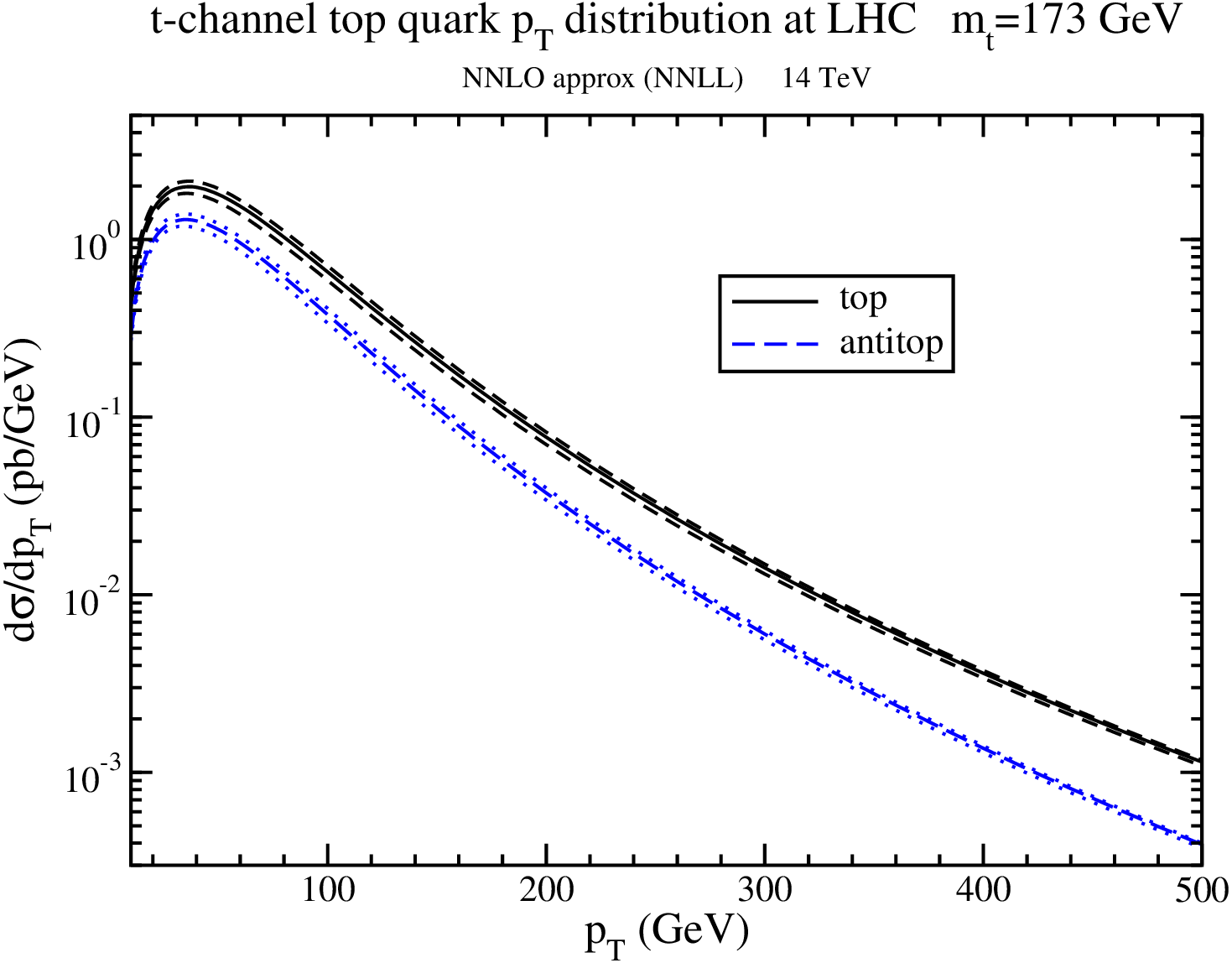}
\caption{NNLO approximate top and antitop 
$p_T$ distributions at 14 TeV energy at the LHC with $p_T$ up to 500 GeV. 
The central results are with $\mu=m_t$ and the 
uncertainty due to scale variation is displayed.}
\label{pttch14lhclogplot}
\end{center}
\end{figure}

Figs. \ref{pttch14lhcplot} and  \ref{pttch14lhclogplot} display the 
corresponding results at 14 TeV LHC energy. At 14 TeV energy, $d\sigma/dp_T$ 
is an order of magnitude larger for a $p_T$ of 500 GeV than at 7 TeV 
energy for both the top and the antitop distributions.

\begin{figure}
\begin{center}
\includegraphics[width=8cm]{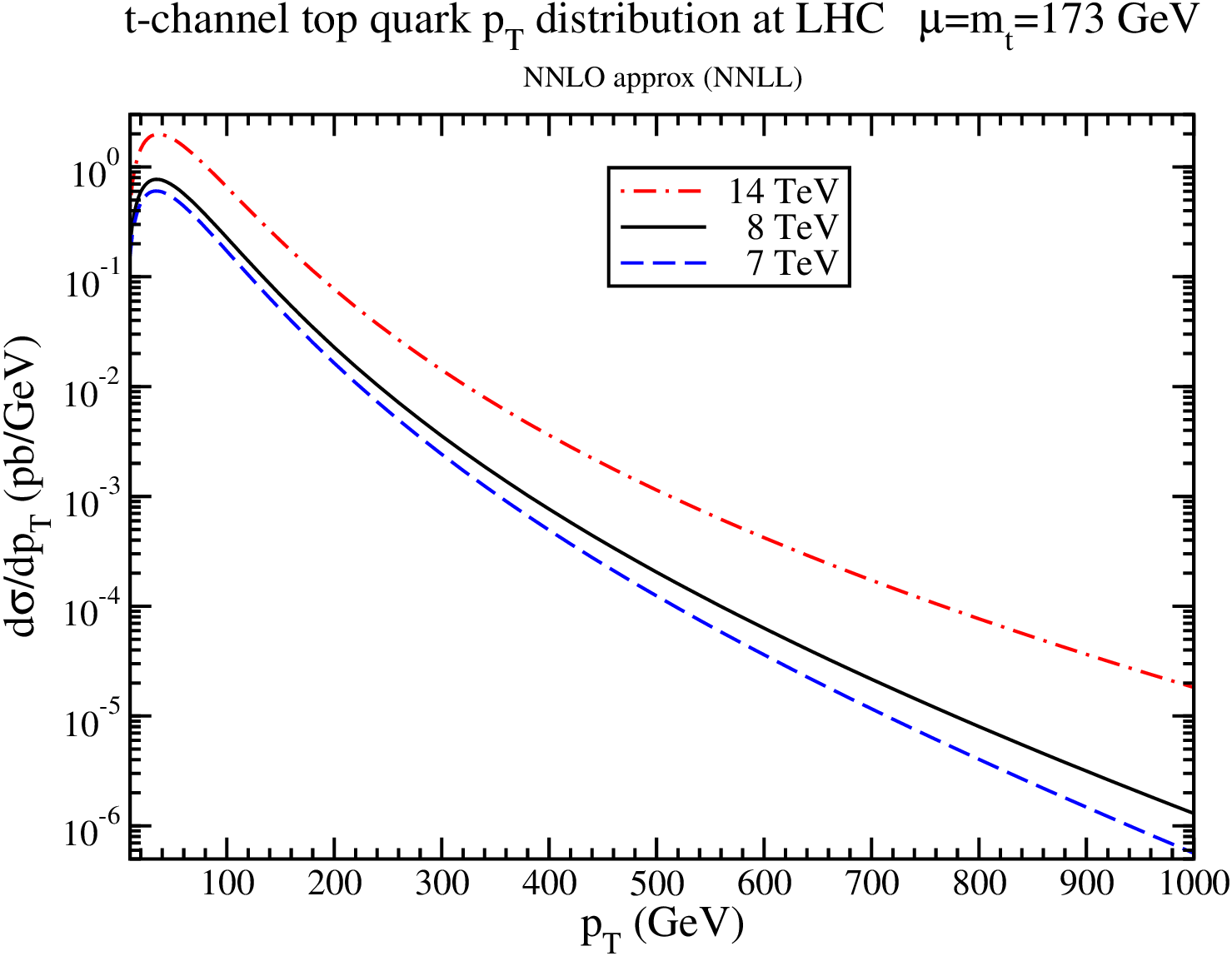}
\hspace{3mm}
\includegraphics[width=8cm]{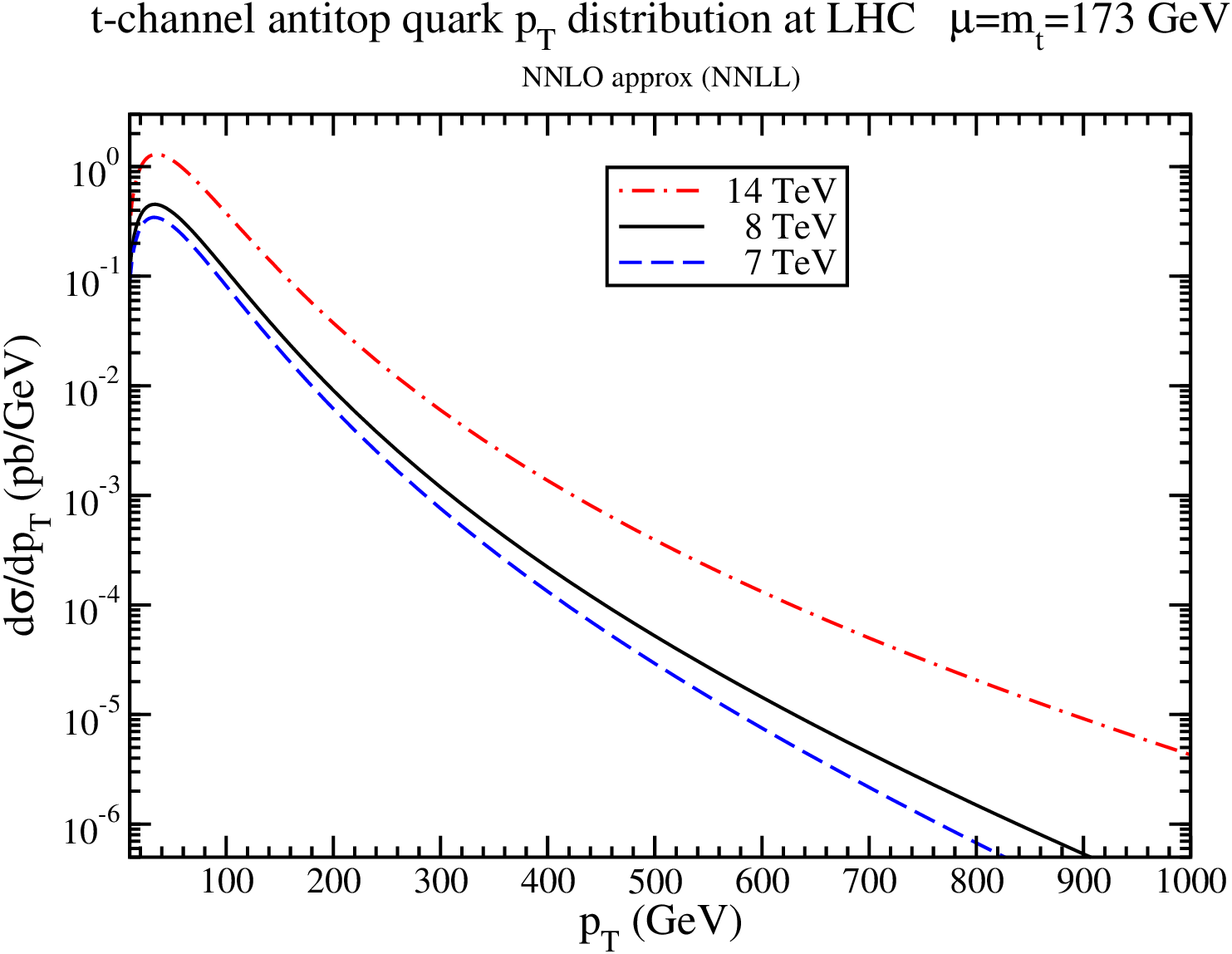}
\caption{Comparison of NNLO approximate top-quark (left) and 
anti-top (right) $p_T$ distributions 
at 7, 8, and 14 TeV energy at the LHC.} 
\label{pttchlhclogplot}
\end{center}
\end{figure}

In Fig. \ref{pttchlhclogplot} we compare the transverse momentum distributions 
for the top (left plot) and the antitop (right plot) at the three different LHC energies of 7, 8, and 14 TeV. The results are all with scale equal to $m_t$ and the $p_T$ range extends up to 1000 GeV. In this $p_T$ range the distributions fall over six orders of magnitude.

\begin{figure}
\begin{center}
\includegraphics[width=8cm]{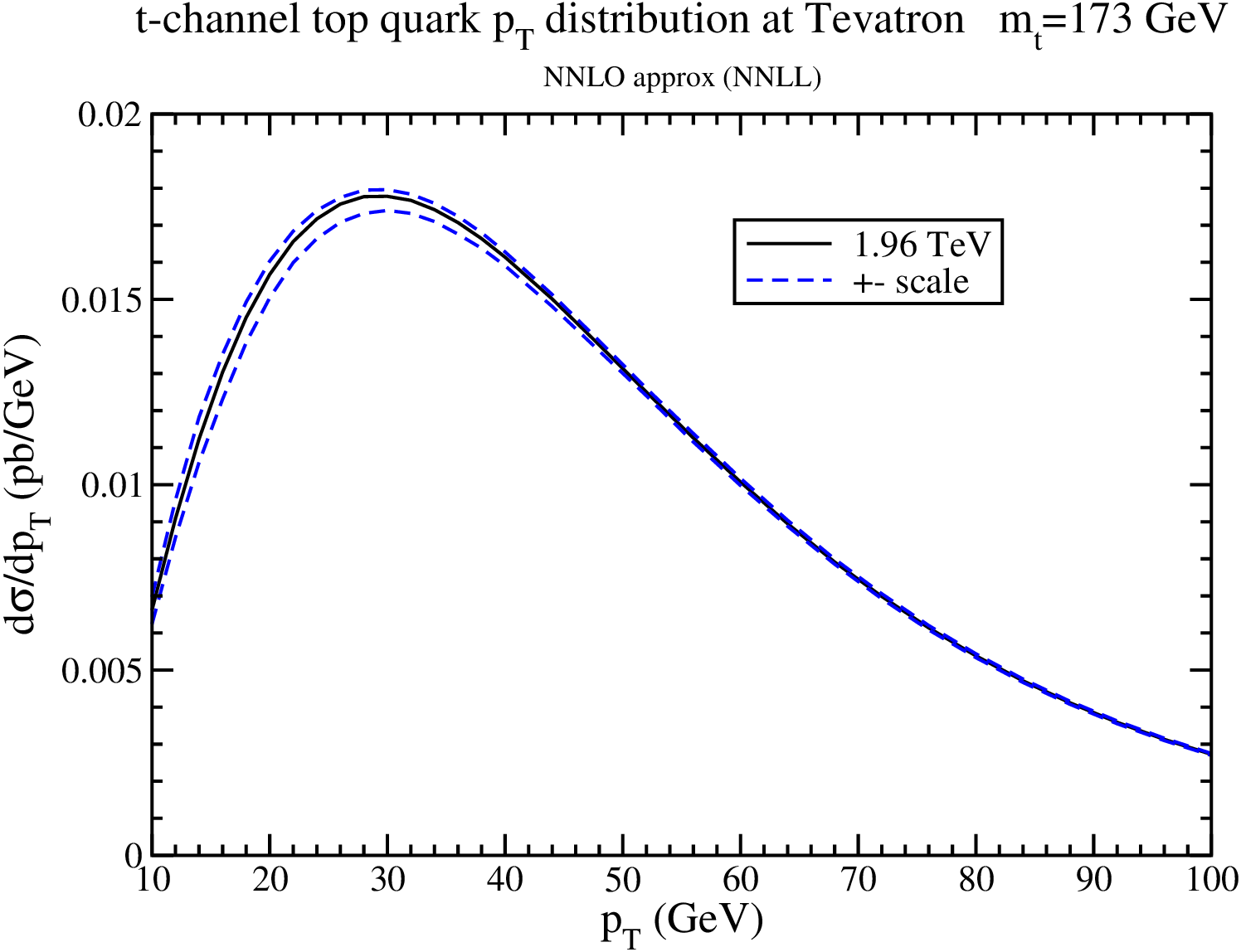}
\hspace{3mm}
\includegraphics[width=8cm]{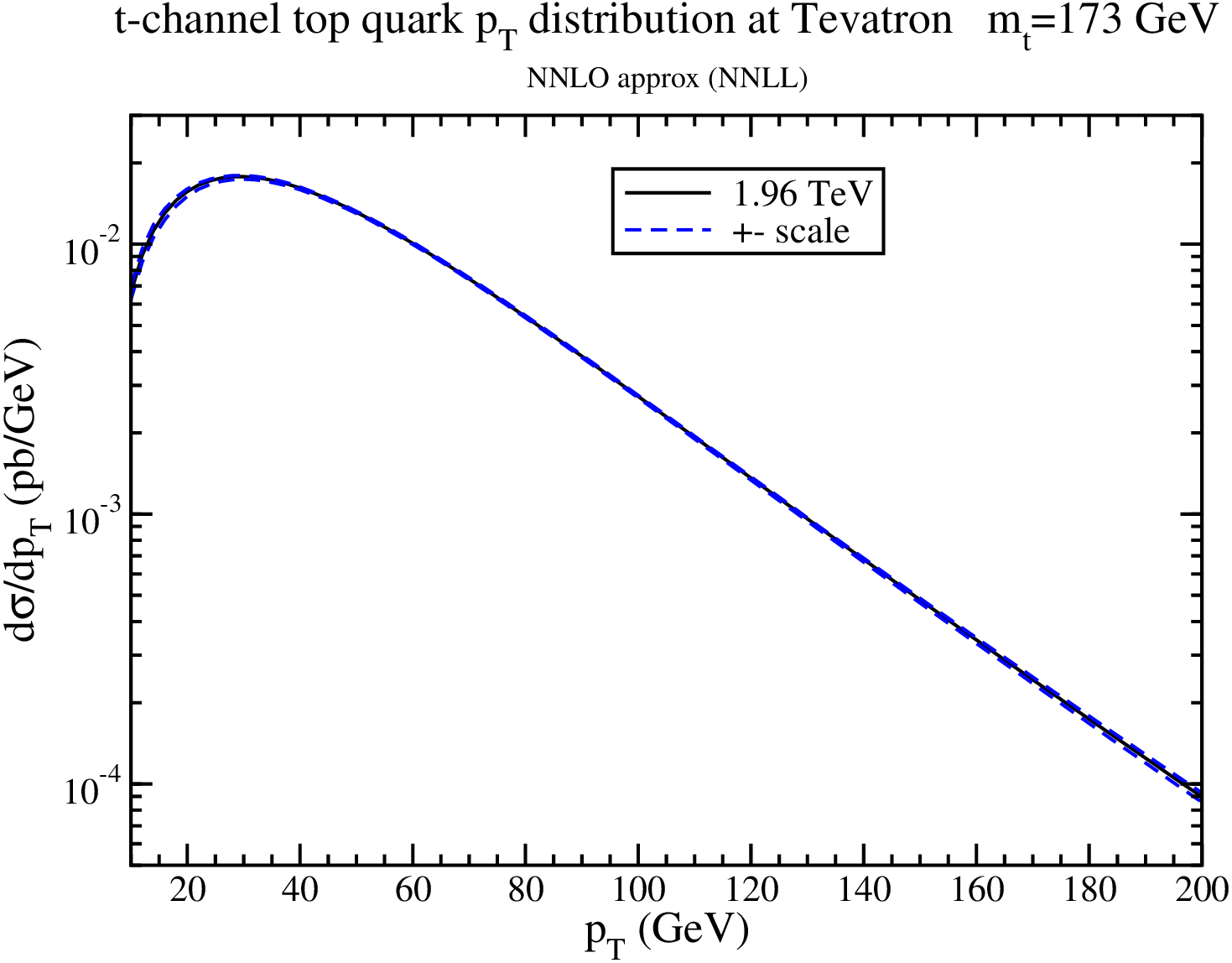}
\caption{NNLO approximate top-quark  
$p_T$ distributions at 1.96 TeV energy at the Tevatron 
in a linear (left) and logarithmic (right) plot. 
The central result is with $\mu=m_t$ and the 
uncertainty due to scale variation is displayed.}
\label{pttchtevplot}
\end{center}
\end{figure}

Finally, in Fig. \ref{pttchtevplot} we present results for the top-quark $p_T$ distribution at the Tevatron with 1.96 TeV energy. The left plot shows results up to a $p_T$ of 100 GeV while the right plot uses a logarithmic scale with results up to a $p_T$ of 200 GeV. The distributions peak at a $p_T$ of around 30 GeV.
The cross section for antitop production at the Tevatron is identical to that for top production.

\mysection{Conclusions}

We have presented the transverse momentum distribution, $d\sigma/dp_T$, 
of the top quark and of the antitop quark in $t$-channel single-top 
and single-antitop production at the LHC and the Tevatron. 
Soft-gluon corrections are known to be important in these processes and 
have been resummed at NNLL accuracy. 
We have improved on NLO calculations by including soft-gluon corrections at 
NNLO. The theoretical 
uncertainty  from factorization and renormalization scale dependence has also 
been determined. The distributions at current LHC energies peak at 
around a $p_T$ of 35 GeV and the NNLO corrections provide an enhancement over 
NLO up to 5\%. 

\mysection*{Acknowledgements}
This material is based upon work supported by the National Science Foundation 
under Grant No. PHY 1212472.


\begin{thebibliography}{99}

\bibitem{D0st}
D0 Collaboration, Phys. Rev. Lett. {\bf 103}, 
092001 (2009) [arXiv:0903.0850 [hep-ex]]; 
Phys. Lett. B {\bf 682}, 363 (2010) [arXiv:0907.4259 [hep-ex]];
Phys. Lett. B {\bf 690}, 5 (2010) [arXiv:0912.1066 [hep-ex]];   
Phys. Lett. B {\bf 705}, 313 (2011) [arXiv:1105.2788 [hep-ex]]; 
Phys. Rev. D {\bf 84}, 112001 (2011) [arXiv:1108.3091 [hep-ex]]. 

\bibitem{CDFst}
CDF Collaboration, Phys. Rev. Lett. {\bf 103}, 
092002 (2009) [arXiv:0903.0885 [hep-ex]]; 
Phys. Rev. D {\bf 82}, 112005 (2010) [arXiv:1004.1181 [hep-ex]]; 
CDF Note 10793.

\bibitem{CMStch}
CMS Collaboration, Phys. Rev. Lett. {\bf 107}, 091802 (2011) 
[arXiv:1106.3052 [hep-ex]];
JHEP 1212 (2012) 035 [arXiv:1209.4533 [hep-ex]]; 
CMS-PAS-TOP-11-021; CMS-PAS-TOP-12-011; CMS-PAS-TOP-12-038.

\bibitem{ATLAStch}
ATLAS Collaboration, Phys. Lett. B {\bf 717}, 330 (2012) 
[arXiv:1205.3130 [hep-ex]];  ATLAS-CONF-2011-088; 
ATLAS-CONF-2011-101; 
ATLAS-CONF-2012-056; ATLAS-CONF-2012-132.

\bibitem{BWHL} 
B.W. Harris, E. Laenen, L. Phaf, Z. Sullivan, and S. Weinzierl,
Phys. Rev. {\bf D 66}, 054024 (2002) [hep-ph/0207055].

\bibitem{CFMT}	
J.M. Campbell, R. Frederix, F. Maltoni, and F. Tramontano, Phys. Rev. Lett. 
{\bf 102}, 182003 (2009) [arXiv:0903.0005 [hep-ph]].
.
\bibitem{SYMC} 
R. Schwienhorst, C.-P. Yuan, C. Mueller, and Q.-H. Cao, Phys. Rev.  D {\bf 83},
034019 (2011) [arXiv:1012.5132 [hep-ph]].

\bibitem{FGMS}	
P. Falgari, F. Giannuzzi, P. Mellor, and A. Signer, 
Phys. Rev. D {\bf 83}, 094013 (2011) [arXiv:1102.5267 [hep-ph]].

\bibitem{FRT}
R. Frederix, E. Re, and P. Torrielli, JHEP 1209 (2012) 130 [arXiv:1207.5391 [hep-ph]].

\bibitem{PF}
P. Falgari, arXiv:1302.3699 [hep-ph].

\bibitem{NKsingletopTev}
N. Kidonakis, Phys. Rev. D {\bf 74}, 114012 (2006) [hep-ph/0609287].

\bibitem{NKsingletopLHC}
N. Kidonakis, Phys. Rev. D {\bf 75}, 071501(R) (2007) [hep-ph/0701080].

\bibitem{NKtch}
N. Kidonakis, Phys. Rev. D {\bf 83}, 091503(R) (2011) [arXiv:1103.2792 [hep-ph]].

\bibitem{NKconf}
N. Kidonakis, in {\sl DIS 2012}, arXiv:1205.3453 [hep-ph]; arXiv:1210.7813 [hep-ph], to appear in a special issue of Particles and Nuclei; in {\sl CKM 2012}, arXiv:1212.2844 [hep-ph].

\bibitem{WLZ}
J. Wang, C.S. Li, and H.X. Zhu, Phys. Rev. D {\bf 87}, 034030 (2013) [arXiv:1210.7698 [hep-ph]]. 

\bibitem{NKBP}
N. Kidonakis and B.D. Pecjak, Eur. Phys. J. C {\bf 72}, 2084 (2012)
[arXiv:1108.6063 [hep-ph]]. 

\bibitem{MSTW}
A.D. Martin, W.J. Stirling, R.S. Thorne, and G. Watt, 
Eur. Phys. J. C {\bf 63}, 189 (2009) [arXiv:0901.0002 [hep-ph]].

\end{thebibliography}
\end{document}